\documentclass[aps,prb,reprint,notitlepage,preprintnumbers,amsmath,amssymb,superscriptaddress, showpacs,showkeys,floatfix]{revtex4-1}

\usepackage{graphicx}
\usepackage{dcolumn}
\usepackage{bm}
\usepackage{color}
\usepackage{braket}
\usepackage[separate-uncertainty,multi-part-units = single ]{siunitx}
\usepackage{lipsum}
\usepackage{float}
\usepackage{bm}
\usepackage{fancyhdr}
\usepackage{romannum}

\usepackage[english]{babel}
\usepackage[utf8]{inputenc}
\usepackage[
	colorlinks=true,
	frenchlinks=false,
        linkcolor=blue,
	    anchorcolor=blue,
        citecolor=blue,
        filecolor=blue,
        urlcolor=blue,
        bookmarks=true,
        bookmarksopen=true,
	    bookmarksnumbered=true,
        bookmarksopenlevel=1,
        plainpages=false,
        pdfpagelabels=true,
	    draft=false]{hyperref}

\begin{document}

\title{Experimental quantum teleportation of propagating microwaves}

\author{K. G. Fedorov}
\email[]{kirill.fedorov@wmi.badw.de}
\affiliation{Walther-Mei{\ss}ner-Institut, Bayerische Akademie der Wissenschaften, 85748 Garching, Germany}
\affiliation{Physik-Department, Technische Universit\"{a}t M\"{u}nchen, 85748 Garching, Germany}

\author{M. Renger}
\affiliation{Walther-Mei{\ss}ner-Institut, Bayerische Akademie der Wissenschaften, 85748 Garching, Germany}
\affiliation{Physik-Department, Technische Universit\"{a}t M\"{u}nchen, 85748 Garching, Germany}

\author{S. Pogorzalek}
\thanks{Present affiliation: IQM, Nymphenburgerstr. 86, 80636 Munich, Germany}
\affiliation{Walther-Mei{\ss}ner-Institut, Bayerische Akademie der Wissenschaften, 85748 Garching, Germany}
\affiliation{Physik-Department, Technische Universit\"{a}t M\"{u}nchen, 85748 Garching, Germany}

\author{R. Di Candia}
\affiliation{Department of Communications and Networking, Aalto University, Espoo, 02150 Finland}

\author{Q. Chen}
\affiliation{Walther-Mei{\ss}ner-Institut, Bayerische Akademie der Wissenschaften, 85748 Garching, Germany}
\affiliation{Physik-Department, Technische Universit\"{a}t M\"{u}nchen, 85748 Garching, Germany}

\author{Y. Nojiri}
\affiliation{Walther-Mei{\ss}ner-Institut, Bayerische Akademie der Wissenschaften, 85748 Garching, Germany}
\affiliation{Physik-Department, Technische Universit\"{a}t M\"{u}nchen, 85748 Garching, Germany}

\author{K. Inomata}
\affiliation{Research Center for Emerging Computing Technologies, Advanced Industrial Science and Technology (AIST), Tsukuba, Ibaraki 305-8568, Japan}
\affiliation{RIKEN Center for Emergent Matter Science (CEMS), Wako, Saitama 351-0198, Japan}

\author{Y. Nakamura}
\affiliation{RIKEN Center for Emergent Matter Science (CEMS), Wako, Saitama 351-0198, Japan}
\affiliation{Research Center for Advanced Science and Technology (RCAST), The University of Tokyo, Meguro-ku, Tokyo 153-8904, Japan}

\author{M. Partanen}
\affiliation{Walther-Mei{\ss}ner-Institut, Bayerische Akademie der Wissenschaften, 85748 Garching, Germany}

\author{A. Marx}
\affiliation{Walther-Mei{\ss}ner-Institut, Bayerische Akademie der Wissenschaften, 85748 Garching, Germany}

\author{R. Gross}
\affiliation{Walther-Mei{\ss}ner-Institut, Bayerische Akademie der Wissenschaften, 85748 Garching, Germany}
\affiliation{Physik-Department, Technische Universit\"{a}t M\"{u}nchen, 85748 Garching, Germany}
\affiliation{Munich Center for Quantum Science and Technology (MCQST), 80799 Munich, Germany}

\author{F. Deppe}
\email[]{frank.deppe@wmi.badw.de}
\affiliation{Walther-Mei{\ss}ner-Institut, Bayerische Akademie der Wissenschaften, 85748 Garching, Germany}
\affiliation{Physik-Department, Technische Universit\"{a}t M\"{u}nchen, 85748 Garching, Germany}
\affiliation{Munich Center for Quantum Science and Technology (MCQST), 80799 Munich, Germany}

\begin{abstract}
The modern field of quantum communication thrives on promise to deliver efficient and unconditionally secure ways to exchange information by exploiting quantum laws of physics. Here, quantum teleportation stands out as an exemplary protocol allowing for the disembodied and safe transfer of unknown quantum states using quantum entanglement and classical communication as resources \cite{Bennett1993}. The experimental feasibility of quantum teleportation with propagating waves, relevant to communication scenarios, has been demonstrated in various physical settings \cite{Bouwmeester1997,Furusawa1998,Pirandola2015}. However, an analogous implementation of quantum teleportation in the microwave domain was missing so far. At the same time, recent breakthroughs in quantum computation with superconducting circuits \cite{Arute2019} have triggered a demand for quantum communication between spatially separated superconducting processors operated at microwave frequencies. Here, we demonstrate a realization of deterministic quantum teleportation of coherent microwave states by exploiting two-mode squeezing and analog feedforward over macroscopic distances $d = 42\,$cm. We achieve teleportation fidelities $F = 0.689 \pm 0.004$ exceeding the no-cloning $F_\mathrm{nc} = 2/3$ threshold for coherent states \cite{Braunstein2001, Pirandola2006} with an average photon number of up to $n_\mathrm{d} = 1.1$. Our results provide a key ingredient for the teleportation-based quantum gate for modular quantum computing with superconducting  circuits and establish a solid foundation for future microwave quantum local area networks \cite{Awschalom2021}.
\end{abstract}

\date{\today}

\maketitle

Quantum teleportation (QT) allows one to achieve a classically impossible goal of transferring an unknown quantum state from one place to another without directly sending it. This task is usually quantified with a teleportation fidelity $F$, which expresses the overlap in the phase space between an unknown input state and a teleported output state. In terms of the corresponding density matrices $\rho_\mathrm{in}$ and $\rho_\mathrm{out}$, the QT fidelity can be expressed as $F = (\mathrm{Tr} \sqrt{\sqrt{\rho_\mathrm{in}} \rho_\mathrm{out} \sqrt{\rho_\mathrm{in}}})^2$. In this case, a transition to the quantum realm occurs when exceeding a so-called classical fidelity threshold $F_\mathrm{ct}$ what is only possible by using nonclassical correlations such as quantum entanglement. The exact value of $F_\mathrm{ct}$ is a subject of many scientific discussions \cite{Braunstein2001}. It depends on the set of teleported states and respective Hilbert space dimension. In the case of continuous-variable propagating fields, $F_\mathrm{ct} = 0$ for arbitrary input states, while for the particular task of teleporting coherent states, one finds $F_\mathrm{ct} = 1/2$ (in contrast to $F_\mathrm{ct} = 2/3$ for qubit states) \cite{Pirandola2006}. This threshold value is connected with a violation of the Clauser-Horne-Shimony-Holt inequality which expresses the fact that nature cannot be described by local hidden-variable theories \cite{Clauser1969, Braunstein2001}.
\begin{figure*}[ht]
        \begin{center}
        \includegraphics[width=1.0\textwidth]{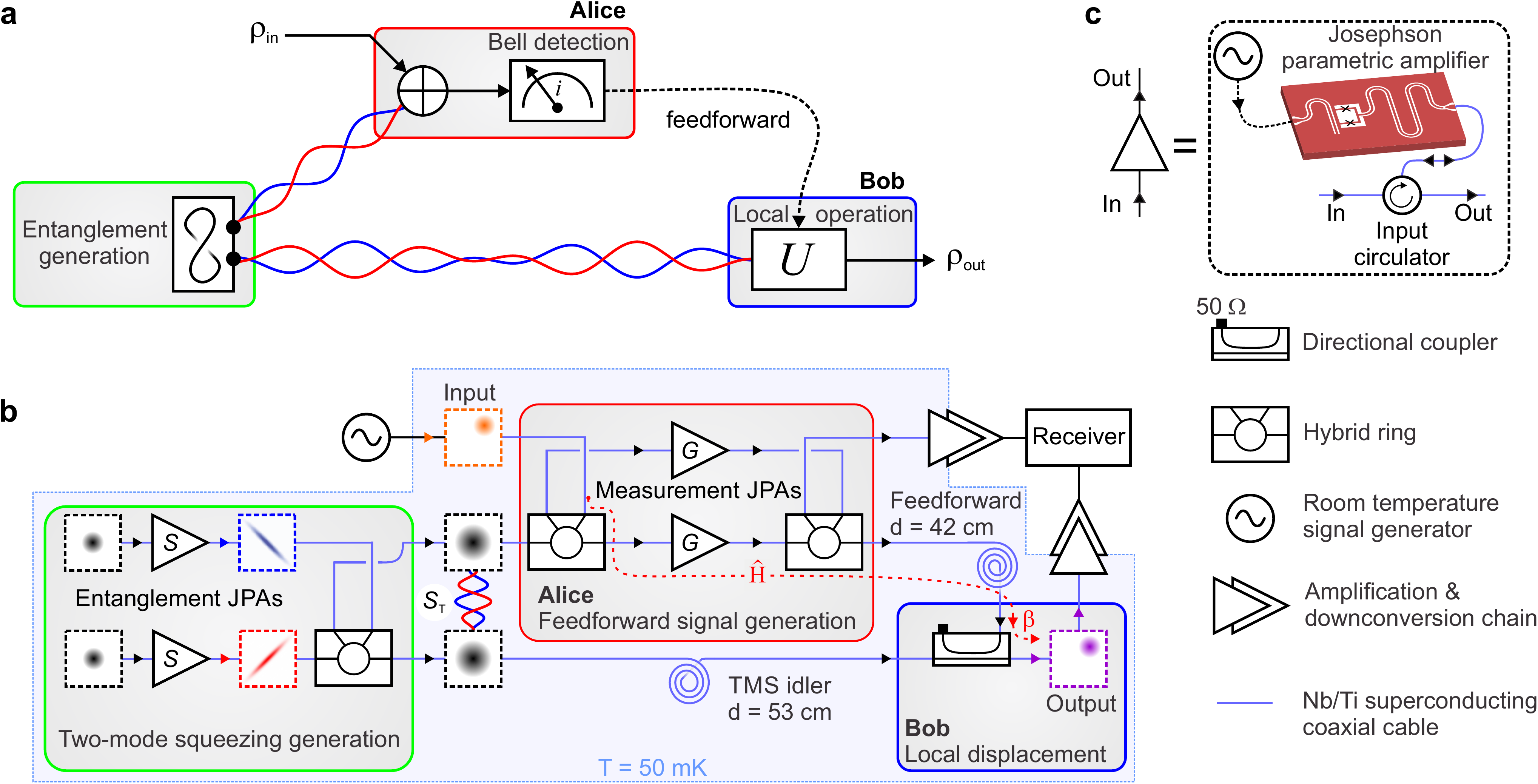}
        \end{center}
    \caption{\textbf{Quantum teleportation of propagating microwaves: concept and implementation.} \textbf{a}, General concept. \textbf{b}, Our experimental implementation of QT with propagating quantum microwaves and analog feedforward. Here, an unknown input coherent state is teleported from Alice to Bob by exploiting quantum entanglement characterized by the two-mode squeezing level $S_\mathrm{T} \lesssim S$. The feedforward signal is generated by the measurement JPAs with the degenerate gain $G$, in combination with two hybrid rings and a local displacement operation on Bob's side. The latter is implemented with a directional coupler with the coupling $\beta = -15$\,dB. Plots in dashed boxes represent quantum states in the quasi-probability Wigner phase space spanned by field quadratures $p$ and $q$. Red dashed line marks a particular input signal path corresponding to operator $\hat{H}$. \textbf{c}, Details and labels of various experimental elements.}
    \label{fig1}
\end{figure*}

A general QT protocol consists of several fundamental steps (also see Fig.\,\ref{fig1}a): (\romannum{1}) entanglement generation and distribution between communication parties (usually denoted as Alice and Bob); (\romannum{2}) local operations on Alice's side aiming at generation of a suitable feedforward signal; (\romannum{3}) feedforward and a local unitary operation on Bob's side. The latter operation results in teleportation of the unknown input state by combining the feedforward signal with the entangled resource state \cite{DiCandia2015}.

Our experimental implementation of QT with propagating microwaves relies on superconducting flux-driven Josephson parametric amplifiers (JPAs) for generation and manipulation of entangled two-mode squeezed microwave states \cite{Yamamoto2008, Menzel2012, Fedorov2018, Pogorzalek2019}. To describe the continuous-variable states of propagating microwave fields, it is convenient to introduce a pair of conjugate operators -- the quadratures $\hat{p}$ and $\hat{q}$. They are analogous to the canonically conjugate variables of position and momentum of a massive particle. A scheme of our QT set-up is shown in Fig.\,\ref{fig1}b. Here, we use two entanglement JPAs in combination with a hybrid ring for generation of path-entangled two-mode squeezed (TMS) states $\ket{\psi_\mathrm{TMS}}$ at the outputs of the hybrid ring \cite{Menzel2012}. These JPAs emit microwave states which are squeezed along different quadratures with the squeezing level  $S$ below vacuum fluctuations. When superimposed at the hybrid ring, these states produce outputs which, locally, look like classical thermal noise. Nevertheless, they possess strong quantum correlations between field quadratures in different propagation paths. These correlations can be described by the two-mode squeezing level $S_\mathrm{T}$, which is, in our case, well-approximated by the local squeezing $S \gtrsim S_\mathrm{T}$ (see Appendix\,\ref{AppA}). In the limit of infinite squeezing $S \rightarrow \infty$, one obtains perfect correlations, e.g. $(\hat{p}_1 + \hat{p}_2) \ket{\psi_\mathrm{TMS}} \rightarrow \delta(p_1 + p_2)$ in the case of anti-correlated $p$-quadratures.

\begin{figure*}[ht]
        \begin{center}
        \includegraphics[width=1.0\textwidth]{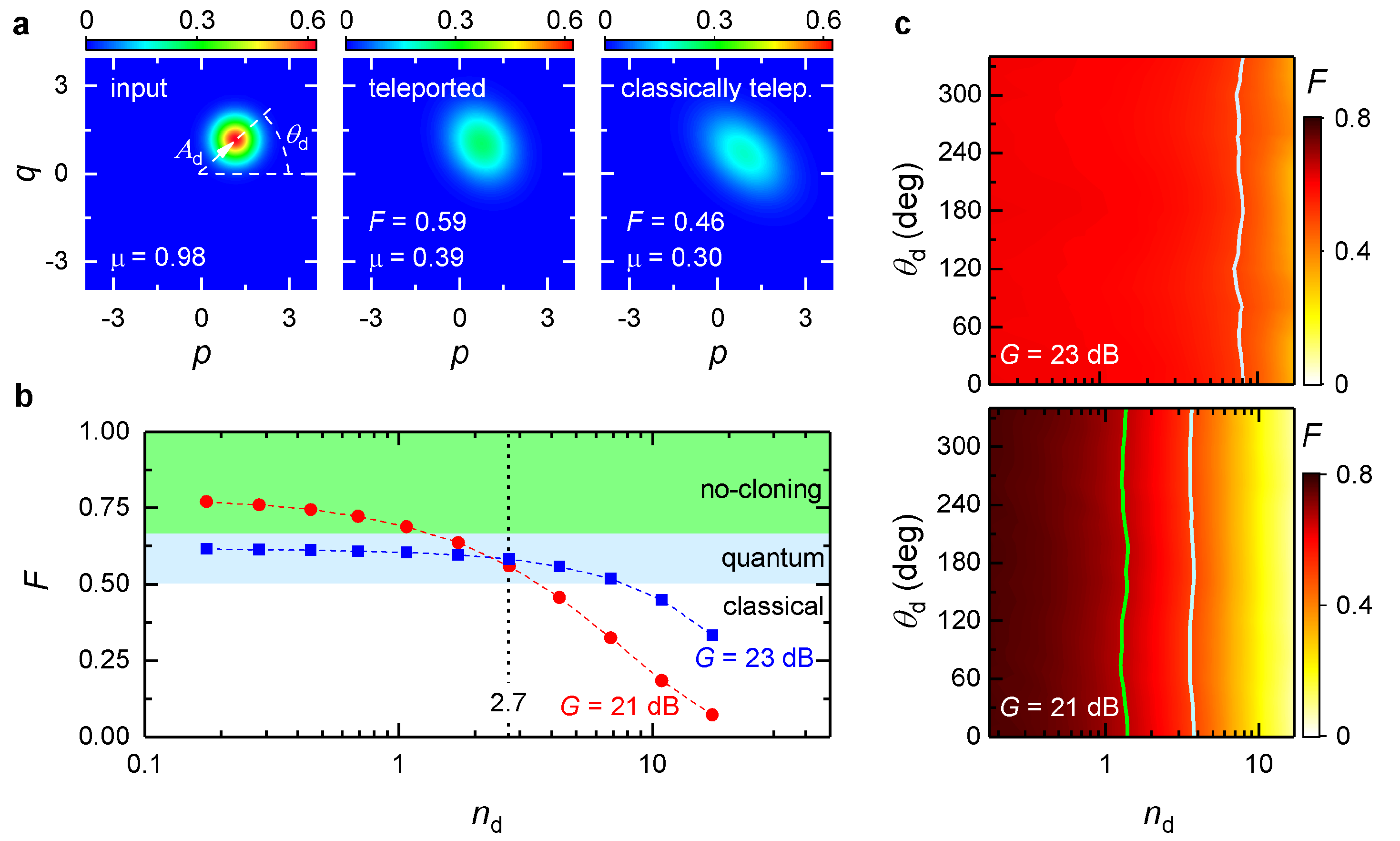}
        \end{center}
    \caption{\textbf{Tomography and fidelity measurements.} \textbf{a}, Reconstructed Wigner functions of an input state, teleported state, and classically teleported state for the squeezing level $S = 4.5$\,dB, the displacement photon number of the input state $n_\mathrm{d} = 2.7$, and the measurement gain $G = 23$\,dB. Inset values represent the quantum teleportation fidelity $F$ and purity $\mu$. \textbf{b}, Fidelity $F$ as a function of $n_\mathrm{d}$ for two characteristic values of $G$. Dashed black line marks the operating point illustrated in panel \textbf{a}. The statistical error is smaller than the symbol size. \textbf{c}, Fidelity $F$ as a function of $n_\mathrm{d}$ and displacement angle $\theta_\mathrm{d}$ for two characteristic values of $G$. Light blue and green lines mark the classical and no-cloning limits, respectively.}
    \label{fig2}
\end{figure*}
Owing to the propagating nature of our TMS states, we can distribute the entangled states between Alice and Bob via superconducting niobium-titanium coaxial cables with characteristic losses as low as 0.001\,dB/m at frequencies $f \simeq 5$\,GHz, thus, implementing step (\romannum{1}) of the QT protocol. On Alice's side, we use another hybrid ring to entangle a weak coherent state, which serves as the unknown input state, with the shared TMS state. The outputs of this second hybrid ring are guided into a pair of measurement JPAs which perform a strong phase-sensitive amplification with the same gain $G$ but along orthogonal amplification angles. By superimposing the outputs of the measurement JPAs at the third hybrid ring, we produce the feedforward signal and conclude step (\romannum{2}). In order to understand the final part of our QT protocol, it is useful to consider a combined action $\hat{H}$ of one of the measurement JPAs with Alice's hybrid rings and Bob's directional coupler with a coupling constant $\beta = -15$\,dB. In the ideal case of zero transmission losses $L = 0$, $\hat{H}$ corresponds to a projective operation when the measurement gain $G$ exactly compensates for the finite coupling $\beta$ and path-losses of two hybrid rings ($3$\,dB each). The corresponding optimal gain is $G_\mathrm{opt} \simeq -\beta + 6\,\mathrm{dB} = 21$\,dB. Indeed, by using the input-output formalism for covariance matrices describing a particular input signal path (depicted by the red dashed line in Fig.\ref{fig1}b), one can show that the combined operator $\hat{H}$ is
\begin{equation}
\hat{H} = \frac{\sqrt{\beta}}{2} \hat{J} \xrightarrow[G \cdot \beta  \to 4]{\beta \to 0} \hat{\Pi}_p =
\begin{pmatrix}
 0 & 0 \\
 0 & 1
\end{pmatrix},
\hat{J} =
\begin{pmatrix}
 1/\sqrt{G} & 0          \\
 0          & \sqrt{G}   \\
\end{pmatrix},
\label{eq1}
\end{equation}
where $\hat{\Pi}_p$ corresponds to a projective measurement of the $p$-quadrature and $\hat{J}$ describes the phase-sensitive amplification by one of the measurement JPAs. Analogously, the other measurement JPA performs projection onto the $q$-quadrature and implements the $\hat{\Pi}_q$ operation. The resulting feedforward application to Bob's part of the TMS state implements teleportation of the input state at the output of the directional coupler, concluding step (\romannum{3}) of the QT protocol (see Supplemental Material \cite{SupMat} for detailed calculations). The demonstrated analog feedforward is fully deterministic and  allows us to tune the strength of our measurement operation by changing the gain $G$ \emph{in situ} and observe its impact on the teleportation protocol. Last but not least, our analog feedforward avoids many issues related to digital feedforward approaches, such as detection inefficiencies or unavoidable temporal delays due to signal processing.

Figure\,\ref{fig2} shows experimental results of the microwave QT protocol. They include Wigner state tomography over a broad range of input coherent states, corresponding quantum-teleported output states, and classically teleported output states (transferred using the same protocol but in the absence of entanglement resource). Here, the carrier frequency of the input, two-mode squeezed, and teleported states is $f = 5.435$\,GHz. Experimental bandwidths of the entanglement and measurement JPAs are $11$\,MHz and $4$\,MHz, respectively. However, the actual measurement bandwidth $\Delta f = 400$\,kHz is limited by a digital finite-impulse response filter in the FPGA-based receiver used for state reconstruction. A particular data set in Fig.\,\ref{fig2}a demonstrates experimental tomography results with the QT fidelity $F = 0.596 \pm 0.004$ larger than the classical threshold $F_\mathrm{ct} = 1/2$. At the same time, classical teleportation yields $F = 0.46 < F_\mathrm{ct}$.

\begin{figure*}[ht]
        \begin{center}
        \includegraphics[width=1.0\textwidth]{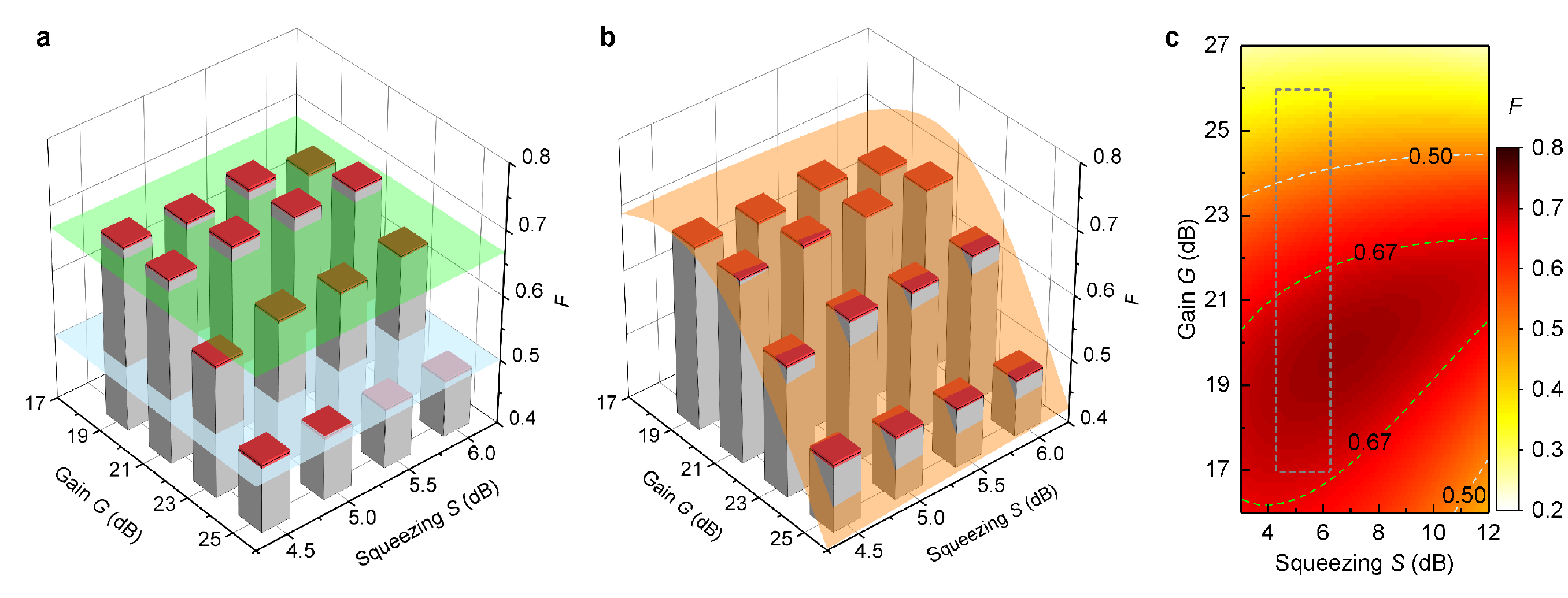}
        \end{center}
    \caption{\textbf{Fidelity thresholds and theory model.} \textbf{a}, Experimental quantum teleportation fidelities $F$ as a function of the measurement gain $G$ and squeezing $S$ for $n_\mathrm{d} = 1.1$ photons. Red bars denote standard deviation of the experimental data. Light blue plane corresponds to the fidelity threshold $F = 0.5$ between quantum and classical regimes, while green plane denotes the no-cloning limit $F_\mathrm{nc} = 2/3$. The experimental data clearly violates the no-cloning limit for $G = 21$\,dB in the whole range of squeezing levels. \textbf{b}, Same data with the fitted theory model (orange plane). \textbf{c}, Extended view over the expected QT performance for the same model, where dark grey dashed box outlines the area presented in panel \textbf{b}. This theory plot demonstrates that further improvement of teleportation fidelities requires an increase of both the measurement gain $G$ and squeezing level $S$.}
    \label{fig3}
\end{figure*}
Strictly speaking, teleportation of a particular coherent state is not sufficient for general purposes of quantum communication. Typically, one needs to demonstrate the successful teleportation of a set of quantum states which could form a communication alphabet (i.e., a codebook). We form this alphabet with a set of weak coherent states by varying both their phase $\theta_\mathrm{d}$ and amplitude $A_\mathrm{d} \sim \sqrt{n_\mathrm{d}}$, where $n_d$ is the displacement photon number. We observe an approximately constant fidelity $F \simeq 0.55$ of the teleported coherent states up to $n_\mathrm{d} \simeq 2$ photons over the whole phase range for $G = 23$\,dB as shown in Fig.\,\ref{fig2}c. Moreover, for the measurement gain $G = 21$\,dB we are also able to violate the no-cloning bound $F_\mathrm{nc} = 2/3$ for $n_\mathrm{d} \leq 1.1$, which is key to the unconditional security of quantum communication \cite{Grosshans2002}. For $n_\mathrm{d} > 1.1$, we observe a smooth degradation of teleportation fidelities, which we attribute to compression effects in the measurement JPAs. In principle, in order to make our QT protocol completely secure against potential eavesdropping of the feedforward signal, we would have to extend the power range of teleported states to $n_\mathrm{d} \gg 1$ while preserving $F > F_\mathrm{nc}$. This goal can be achieved in the future by improving the 1-dB compression point of our JPAs, $P_{\mathrm{1}\mbox{-}\mathrm{dB}} \simeq -130$\,dBm, to higher values \cite{Sivak2020}. A weak dependence of the QT fidelity $F$ on the coherent phase $\theta_\mathrm{d}$ , observable for $n_\mathrm{d} \gg 1$ in Fig.\,\ref{fig2}c, arises from an interplay between the compression effects and non-perfect orthogonality of phase-sensitive amplification in the measurement JPAs.

Finally, we investigate the influence of both the measurement JPA gain $G$ and entanglement strength expressed via the squeezing level $S$ on the QT performance. This task is straightforwardly achieved in our experimental setting by tuning microwave pump amplitudes of the measurement and entanglement JPAs. Figure\,\ref{fig3}a shows the experimental fidelity $F$ of coherent state teleportation as a function of $G$ and $S$. One can immediately notice a fidelity maximum at $G = 21$\,dB as naively expected from theory. However, one has to keep in mind that $G_\mathrm{opt} = 21$\,dB is only valid in the limit of $S \to \infty$ and $L = 0$. In reality, our squeezing levels are far from infinity. Also, we estimate microwave losses $L \simeq 2$\,dB due to reflections and dissipation in non-superconducting parts. In order to describe these non-idealities, we develop a theory model based on the input-output formalism and use it to fit the experimental data. This fit has three fitting parameters: temperature $T$ of the electromagnetic environment, and noise parameters $\chi_1$ and $\chi_2$ which describe the JPA gain dependence of the noise photon number $n = \chi_1 (G - 1)^{\chi_2}$. Here, $n$ is referred to the JPA input and $G$ corresponds to the JPA degenerate gain \cite{Renger2020}. All other model parameters, such as various losses, are determined from independent measurements and can be found in Supplemental Material \cite{SupMat}. We find a good qualitative agreement between theory and experiment, as it can be seen in Fig.\,\ref{fig3}b. Our model also demonstrates that the optimal measurement gain is weakly dependent on $S$ and becomes $G_\mathrm{opt}^{'} = 23$\,dB for $S \to \infty$. This can be intuitively understood by considering the fact that an asymptotic optimal gain also needs to compensate for finite transmission losses $G_\mathrm{opt}^{'} = G_\mathrm{opt} + L$, while $G = 21$\,dB appears to be only a local optimum for the finite squeezing levels accessible in our experiments. This is also in agreement with the fact that the optimal gain increases with the increasing entanglement strength \cite{Scorpo2017}.

To quantify the amount of information, which can be securely sent via quantum teleportation of coherent states, we utilize the effective number of classical bits encoded in one of the teleported state parameters. Here, we focus on the coherent state phase $\theta_\mathrm{d}$, since it is most suitable for communication of classical information for a fixed $n_\mathrm{d}$. We estimate the corresponding bit rate $N$ in our QT protocol by using the well-known Shannon-Hartley equation for the channel capacity, $N = \Delta f \log_2 (1 + \mathrm{SNR})$. In our case, the signal-to-noise ratio $\mathrm{SNR}$ can be expressed as $\mathrm{SNR} =  \Delta\theta_\mathrm{d}^2 / \sigma^2$, where $\Delta\theta_\mathrm{d}$ is the available range of coherent phases and $\sigma$ is the standard deviation of $\theta_\mathrm{d}$. In order to obtain a lower bound for $\sigma$, we use the quantum Cram\'{e}r-Rao bound \cite{Braunstein1994}, which relates the quantum Fisher information $\mathcal{F}_\mathrm{Q}$ to a lower bound of the estimation error of $\theta_\mathrm{d}$ as $\sigma^2 \geq 1 / \mathcal{F}_\mathrm{Q}$. This bound can be saturated by a suitable measurement choice. The quantum Fisher information $\mathcal{F}_\mathrm{Q}$ for phase estimation of a displaced Gaussian state is given by \cite{Pinel2013}
\begin{equation}
\mathcal{F}_\mathrm{Q} = 4 \mu n_\mathrm{d} (\lambda \cos^2\phi + \lambda^{-1} \sin^2\phi) + \frac{1}{1+\mu^2} \frac{(1-\lambda^2)^2}{\lambda^2},
\label{QFisher}
\end{equation}
where $\lambda\,=\,\exp(-2 r)$ is the squeezed variance, $\phi$ is the squeezing angle, and $\mu\,=\,\mathrm{Tr}(\rho_\mathrm{out}^2)$ is the state purity. Here, all parameters are those of the teleported states. By using (\ref{QFisher}) for the experimental QT parameters corresponding to the working point with $G = 21$\,dB, $S = 6$\,dB, and $n_\mathrm{d} = 1.1$, we obtain $\mathcal{F}_\mathrm{Q} \simeq 0.85$. From Fig.\,\ref{fig2}, we see that $\Delta\theta_\mathrm{d} = 2 \pi$, which results in the upper bound for the achievable bit-rate $N = 1.95$\,Mbits/s.

In conclusion, we have demonstrated a successful experimental implementation of the quantum teleportation protocol with propagating microwaves over a distance of 42\,cm in the cryogenic environment. Our teleportation protocol allows us to violate the no-cloning limit over a wide range of input state parameters, corresponding to teleportation fidelities of $F \geq 0.69$ for coherent states with the displacement photon number of $n_\mathrm{d} \leq 1.1$ and the measurement gain $G = 21$\,dB. Our experimental techniques rely exclusively on conventional aluminum--niobium superconducting parametric devices for generation and control of quantum microwave signals, which makes them fully compatible with other quantum superconducting circuits in terms of frequencies, such as microwave quantum memory cells \cite{Edwar2018, Gao2018}, and fabrication technology. This natural technology matching also avoids massive ($\sim 10^{-5}$) conversion losses of state-of-the-art transducers between optical and microwave frequencies at the single photon level \cite{Mirhosseini2020}. We envision our results to be useful for creating quantum local area networks between superconducting quantum computers. Our experiments pave the road towards distributed superconducting quantum supercomputers in the future and allow one to exploit advantages of secure quantum communication in the convenient microwave regime.

\appendix
\section{}
\label{AppA}

The task of the entanglement JPAs is to perform a local squeezing operation $\hat{S}(\xi) |0\rangle$, which can be described by a single-mode squeezing operator $\hat{S}(\xi) = \exp(\frac{1}{2} \xi ^* \hat{a}^2 \,{-}\, \frac{1}{2}\xi (\hat{a}^\dagger)^2)$, where $\hat{a}^{\dagger} = \hat{q}-i\hat{p}$ and $\hat{a}\,{=}\,\hat{q}+i\hat{p}$ are the creation and annihilation operators with $\left[\hat{a},\hat{a}^\dagger\right]\,{=}\,1$ of the $f_0$ mode with quadratures $\hat{q}$ and $\hat{p}$, and $\xi = r e^{i\phi}$ is the complex squeezing amplitude.  Here, the phase $\phi \,{=-}2\gamma$ determines the squeezing angle $\gamma$ between the antisqueezed quadrature and the $p$-axis in the phase space, while the squeezing factor $r$ parameterizes the amount of squeezing. We define the degree of squeezing in decibels as $S \,{=}\,{-}10\,\log_{10} (\sigma_\mathrm{s}^2 / 0.25)$, where $\sigma_\mathrm{s}^2$ is the variance of the squeezed quadrature and the vacuum variance is $0.25$. Positive values of $S$ indicate squeezing below the vacuum level. The antisqueezing level is defined as $A\,{=}\,10\,\log_{10} (\sigma_\mathrm{a}^2 / 0.25)$, where $\sigma_\mathrm{a}^2$ is the variance of the antisqueezed quadrature. We generate symmetric two-mode squeezed (TMS) states at the output of the hybrid ring by pumping entanglement JPAs\,1 and 2 with strong quasi-continuous microwave drives so that they produce squeezed vacuum states with the same squeezing level but orthogonal squeezing angles $\gamma_2\,{=}\,\gamma_1\,{+}\,\pi/2$. These angles are stabilized by controlling the respective pump phases employing a phase-locked loop\cite{Fedorov2016,Fedorov2018}. A TMS state can be described by the two-mode squeezing operator $\hat{S}_\mathrm{T}=\exp\left(\xi_\mathrm{T}^* \hat{a}_1\hat{a}_2 - \xi_\mathrm{T} \hat{a}^\dagger_1\hat{a}^\dagger_2\right)$, where $\hat{a}_i$ is the annihilation operator of the $i$-th electromagnetic mode and $\xi_\mathrm{T} = r_\mathrm{T} e^{i \varphi}$. Here, the amount of two-mode squeezing is given by $S_\mathrm{T} = -10\,\log_{10} (- 2 r_\mathrm{T})$ and the phase $\varphi$ determines which quadratures on the two modes are correlated. The difference between the genuine two-mode squeezing level $S_\mathrm{T}$ (at the hybrid ring outputs) and local squeezing level $S$ (at the hybrid ring inputs) is fully defined by the hybrid ring insertion losses $L_\mathrm{HR} = 0.4\,$dB and environmental noise photon number $n_\mathrm{env} \simeq 0.025$. In our case, this amounts to a small difference of around $10\,\%$. Thus, we can use $S$ as a good direct quantifier for the amount of two-mode squeezing in the propagating microwave signals.

In order to reconstruct the quantum states in the experiment, we employ a well-tested reference state tomography based on statistical moments of the detected field quadratures\cite{Menzel2012, Eichler2011b}.

\medskip\noindent
\textbf{\large Acknowledgments}

\noindent
We acknowledge support by the German Research Foundation via Germany’s Excellence Strategy (EXC-2111-390814868), Elite Network of Bavaria through the program ExQM, EU Flagship project QMiCS (Grant No. 820505), the Marie Skłodowska Curie fellowship number 891517 (MSC-IF Green-MIQUEC), and JST ERATO (Grant No. JPMJER1601).

\bibliography{QT_bib.bib}

\begin{thebibliography}{26}%
\makeatletter
\providecommand \@ifxundefined [1]{%
 \@ifx{#1\undefined}
}%
\providecommand \@ifnum [1]{%
 \ifnum #1\expandafter \@firstoftwo
 \else \expandafter \@secondoftwo
 \fi
}%
\providecommand \@ifx [1]{%
 \ifx #1\expandafter \@firstoftwo
 \else \expandafter \@secondoftwo
 \fi
}%
\providecommand \natexlab [1]{#1}%
\providecommand \enquote  [1]{``#1''}%
\providecommand \bibnamefont  [1]{#1}%
\providecommand \bibfnamefont [1]{#1}%
\providecommand \citenamefont [1]{#1}%
\providecommand \href@noop [0]{\@secondoftwo}%
\providecommand \href [0]{\begingroup \@sanitize@url \@href}%
\providecommand \@href[1]{\@@startlink{#1}\@@href}%
\providecommand \@@href[1]{\endgroup#1\@@endlink}%
\providecommand \@sanitize@url [0]{\catcode `\\12\catcode `\$12\catcode
  `\&12\catcode `\#12\catcode `\^12\catcode `\_12\catcode `\%12\relax}%
\providecommand \@@startlink[1]{}%
\providecommand \@@endlink[0]{}%
\providecommand \url  [0]{\begingroup\@sanitize@url \@url }%
\providecommand \@url [1]{\endgroup\@href {#1}{\urlprefix }}%
\providecommand \urlprefix  [0]{URL }%
\providecommand \Eprint [0]{\href }%
\providecommand \doibase [0]{http://dx.doi.org/}%
\providecommand \selectlanguage [0]{\@gobble}%
\providecommand \bibinfo  [0]{\@secondoftwo}%
\providecommand \bibfield  [0]{\@secondoftwo}%
\providecommand \translation [1]{[#1]}%
\providecommand \BibitemOpen [0]{}%
\providecommand \bibitemStop [0]{}%
\providecommand \bibitemNoStop [0]{.\EOS\space}%
\providecommand \EOS [0]{\spacefactor3000\relax}%
\providecommand \BibitemShut  [1]{\csname bibitem#1\endcsname}%
\let\auto@bib@innerbib\@empty
\bibitem [{\citenamefont {Bennett}\ \emph {et~al.}(1993)\citenamefont
  {Bennett}, \citenamefont {Brassard}, \citenamefont {Cr\'epeau}, \citenamefont
  {Jozsa}, \citenamefont {Peres},\ and\ \citenamefont
  {Wootters}}]{Bennett1993}%
  \BibitemOpen
  \bibfield  {author} {\bibinfo {author} {\bibfnamefont {C.~H.}\ \bibnamefont
  {Bennett}}, \bibinfo {author} {\bibfnamefont {G.}~\bibnamefont {Brassard}},
  \bibinfo {author} {\bibfnamefont {C.}~\bibnamefont {Cr\'epeau}}, \bibinfo
  {author} {\bibfnamefont {R.}~\bibnamefont {Jozsa}}, \bibinfo {author}
  {\bibfnamefont {A.}~\bibnamefont {Peres}}, \ and\ \bibinfo {author}
  {\bibfnamefont {W.~K.}\ \bibnamefont {Wootters}},\ }\href {\doibase
  10.1103/PhysRevLett.70.1895} {\bibfield  {journal} {\bibinfo  {journal}
  {Phys. Rev. Lett.}\ }\textbf {\bibinfo {volume} {70}},\ \bibinfo {pages}
  {1895} (\bibinfo {year} {1993})}\BibitemShut {NoStop}%
\bibitem [{\citenamefont {Bouwmeester}\ \emph {et~al.}(1997)\citenamefont
  {Bouwmeester}, \citenamefont {Pan}, \citenamefont {Mattle}, \citenamefont
  {Eibl}, \citenamefont {Weinfurter},\ and\ \citenamefont
  {Zeilinger}}]{Bouwmeester1997}%
  \BibitemOpen
  \bibfield  {author} {\bibinfo {author} {\bibfnamefont {D.}~\bibnamefont
  {Bouwmeester}}, \bibinfo {author} {\bibfnamefont {J.-W.}\ \bibnamefont
  {Pan}}, \bibinfo {author} {\bibfnamefont {K.}~\bibnamefont {Mattle}},
  \bibinfo {author} {\bibfnamefont {M.}~\bibnamefont {Eibl}}, \bibinfo {author}
  {\bibfnamefont {H.}~\bibnamefont {Weinfurter}}, \ and\ \bibinfo {author}
  {\bibfnamefont {A.}~\bibnamefont {Zeilinger}},\ }\href {\doibase
  10.1038/37539} {\bibfield  {journal} {\bibinfo  {journal} {Nature}\ }\textbf
  {\bibinfo {volume} {390}},\ \bibinfo {pages} {575} (\bibinfo {year}
  {1997})}\BibitemShut {NoStop}%
\bibitem [{\citenamefont {Furusawa}\ \emph {et~al.}(1998)\citenamefont
  {Furusawa}, \citenamefont {S{\o}rensen}, \citenamefont {Braunstein},
  \citenamefont {Fuchs}, \citenamefont {Kimble},\ and\ \citenamefont
  {Polzik}}]{Furusawa1998}%
  \BibitemOpen
  \bibfield  {author} {\bibinfo {author} {\bibfnamefont {A.}~\bibnamefont
  {Furusawa}}, \bibinfo {author} {\bibfnamefont {J.~L.}\ \bibnamefont
  {S{\o}rensen}}, \bibinfo {author} {\bibfnamefont {S.~L.}\ \bibnamefont
  {Braunstein}}, \bibinfo {author} {\bibfnamefont {C.~A.}\ \bibnamefont
  {Fuchs}}, \bibinfo {author} {\bibfnamefont {H.~J.}\ \bibnamefont {Kimble}}, \
  and\ \bibinfo {author} {\bibfnamefont {E.~S.}\ \bibnamefont {Polzik}},\
  }\href {\doibase 10.1126/science.282.5389.706} {\bibfield  {journal}
  {\bibinfo  {journal} {Science}\ }\textbf {\bibinfo {volume} {282}},\ \bibinfo
  {pages} {706} (\bibinfo {year} {1998})}\BibitemShut {NoStop}%
\bibitem [{\citenamefont {Pirandola}\ \emph {et~al.}(2015)\citenamefont
  {Pirandola}, \citenamefont {Eisert}, \citenamefont {Weedbrook}, \citenamefont
  {Furusawa},\ and\ \citenamefont {Braunstein}}]{Pirandola2015}%
  \BibitemOpen
  \bibfield  {author} {\bibinfo {author} {\bibfnamefont {S.}~\bibnamefont
  {Pirandola}}, \bibinfo {author} {\bibfnamefont {J.}~\bibnamefont {Eisert}},
  \bibinfo {author} {\bibfnamefont {C.}~\bibnamefont {Weedbrook}}, \bibinfo
  {author} {\bibfnamefont {A.}~\bibnamefont {Furusawa}}, \ and\ \bibinfo
  {author} {\bibfnamefont {S.~L.}\ \bibnamefont {Braunstein}},\ }\href
  {\doibase 10.1038/nphoton.2015.154} {\bibfield  {journal} {\bibinfo
  {journal} {Nat. Photon.}\ }\textbf {\bibinfo {volume} {9}},\ \bibinfo {pages}
  {641} (\bibinfo {year} {2015})}\BibitemShut {NoStop}%
\bibitem [{\citenamefont {Arute}\ \emph {et~al.}(2019)\citenamefont {Arute},
  \citenamefont {Arya}, \citenamefont {Babbush}, \citenamefont {Bacon},
  \citenamefont {Bardin}, \citenamefont {Barends}, \citenamefont {Biswas},
  \citenamefont {Boixo}, \citenamefont {Brandao}, \citenamefont {Buell},
  \citenamefont {Burkett}, \citenamefont {Chen}, \citenamefont {Chen},
  \citenamefont {Chiaro}, \citenamefont {Collins}, \citenamefont {Courtney},
  \citenamefont {Dunsworth}, \citenamefont {Farhi}, \citenamefont {Foxen},
  \citenamefont {Fowler}, \citenamefont {Gidney}, \citenamefont {Giustina},
  \citenamefont {Graff}, \citenamefont {Guerin}, \citenamefont {Habegger},
  \citenamefont {Harrigan}, \citenamefont {Hartmann}, \citenamefont {Ho},
  \citenamefont {Hoffmann}, \citenamefont {Huang}, \citenamefont {Humble},
  \citenamefont {Isakov}, \citenamefont {Jeffrey}, \citenamefont {Jiang},
  \citenamefont {Kafri}, \citenamefont {Kechedzhi}, \citenamefont {Kelly},
  \citenamefont {Klimov}, \citenamefont {Knysh}, \citenamefont {Korotkov},
  \citenamefont {Kostritsa}, \citenamefont {Landhuis}, \citenamefont
  {Lindmark}, \citenamefont {Lucero}, \citenamefont {Lyakh}, \citenamefont
  {Mandr{\`a}}, \citenamefont {McClean}, \citenamefont {McEwen}, \citenamefont
  {Megrant}, \citenamefont {Mi}, \citenamefont {Michielsen}, \citenamefont
  {Mohseni}, \citenamefont {Mutus}, \citenamefont {Naaman}, \citenamefont
  {Neeley}, \citenamefont {Neill}, \citenamefont {Niu}, \citenamefont {Ostby},
  \citenamefont {Petukhov}, \citenamefont {Platt}, \citenamefont {Quintana},
  \citenamefont {Rieffel}, \citenamefont {Roushan}, \citenamefont {Rubin},
  \citenamefont {Sank}, \citenamefont {Satzinger}, \citenamefont {Smelyanskiy},
  \citenamefont {Sung}, \citenamefont {Trevithick}, \citenamefont
  {Vainsencher}, \citenamefont {Villalonga}, \citenamefont {White},
  \citenamefont {Yao}, \citenamefont {Yeh}, \citenamefont {Zalcman},
  \citenamefont {Neven},\ and\ \citenamefont {Martinis}}]{Arute2019}%
  \BibitemOpen
  \bibfield  {author} {\bibinfo {author} {\bibfnamefont {F.}~\bibnamefont
  {Arute}}, \bibinfo {author} {\bibfnamefont {K.}~\bibnamefont {Arya}},
  \bibinfo {author} {\bibfnamefont {R.}~\bibnamefont {Babbush}}, \bibinfo
  {author} {\bibfnamefont {D.}~\bibnamefont {Bacon}}, \bibinfo {author}
  {\bibfnamefont {J.~C.}\ \bibnamefont {Bardin}}, \bibinfo {author}
  {\bibfnamefont {R.}~\bibnamefont {Barends}}, \bibinfo {author} {\bibfnamefont
  {R.}~\bibnamefont {Biswas}}, \bibinfo {author} {\bibfnamefont
  {S.}~\bibnamefont {Boixo}}, \bibinfo {author} {\bibfnamefont {F.~G. S.~L.}\
  \bibnamefont {Brandao}}, \bibinfo {author} {\bibfnamefont {D.~A.}\
  \bibnamefont {Buell}}, \bibinfo {author} {\bibfnamefont {B.}~\bibnamefont
  {Burkett}}, \bibinfo {author} {\bibfnamefont {Y.}~\bibnamefont {Chen}},
  \bibinfo {author} {\bibfnamefont {Z.}~\bibnamefont {Chen}}, \bibinfo {author}
  {\bibfnamefont {B.}~\bibnamefont {Chiaro}}, \bibinfo {author} {\bibfnamefont
  {R.}~\bibnamefont {Collins}}, \bibinfo {author} {\bibfnamefont
  {W.}~\bibnamefont {Courtney}}, \bibinfo {author} {\bibfnamefont
  {A.}~\bibnamefont {Dunsworth}}, \bibinfo {author} {\bibfnamefont
  {E.}~\bibnamefont {Farhi}}, \bibinfo {author} {\bibfnamefont
  {B.}~\bibnamefont {Foxen}}, \bibinfo {author} {\bibfnamefont
  {A.}~\bibnamefont {Fowler}}, \bibinfo {author} {\bibfnamefont
  {C.}~\bibnamefont {Gidney}}, \bibinfo {author} {\bibfnamefont
  {M.}~\bibnamefont {Giustina}}, \bibinfo {author} {\bibfnamefont
  {R.}~\bibnamefont {Graff}}, \bibinfo {author} {\bibfnamefont
  {K.}~\bibnamefont {Guerin}}, \bibinfo {author} {\bibfnamefont
  {S.}~\bibnamefont {Habegger}}, \bibinfo {author} {\bibfnamefont {M.~P.}\
  \bibnamefont {Harrigan}}, \bibinfo {author} {\bibfnamefont {M.~J.}\
  \bibnamefont {Hartmann}}, \bibinfo {author} {\bibfnamefont {A.}~\bibnamefont
  {Ho}}, \bibinfo {author} {\bibfnamefont {M.}~\bibnamefont {Hoffmann}},
  \bibinfo {author} {\bibfnamefont {T.}~\bibnamefont {Huang}}, \bibinfo
  {author} {\bibfnamefont {T.~S.}\ \bibnamefont {Humble}}, \bibinfo {author}
  {\bibfnamefont {S.~V.}\ \bibnamefont {Isakov}}, \bibinfo {author}
  {\bibfnamefont {E.}~\bibnamefont {Jeffrey}}, \bibinfo {author} {\bibfnamefont
  {Z.}~\bibnamefont {Jiang}}, \bibinfo {author} {\bibfnamefont
  {D.}~\bibnamefont {Kafri}}, \bibinfo {author} {\bibfnamefont
  {K.}~\bibnamefont {Kechedzhi}}, \bibinfo {author} {\bibfnamefont
  {J.}~\bibnamefont {Kelly}}, \bibinfo {author} {\bibfnamefont {P.~V.}\
  \bibnamefont {Klimov}}, \bibinfo {author} {\bibfnamefont {S.}~\bibnamefont
  {Knysh}}, \bibinfo {author} {\bibfnamefont {A.}~\bibnamefont {Korotkov}},
  \bibinfo {author} {\bibfnamefont {F.}~\bibnamefont {Kostritsa}}, \bibinfo
  {author} {\bibfnamefont {D.}~\bibnamefont {Landhuis}}, \bibinfo {author}
  {\bibfnamefont {M.}~\bibnamefont {Lindmark}}, \bibinfo {author}
  {\bibfnamefont {E.}~\bibnamefont {Lucero}}, \bibinfo {author} {\bibfnamefont
  {D.}~\bibnamefont {Lyakh}}, \bibinfo {author} {\bibfnamefont
  {S.}~\bibnamefont {Mandr{\`a}}}, \bibinfo {author} {\bibfnamefont {J.~R.}\
  \bibnamefont {McClean}}, \bibinfo {author} {\bibfnamefont {M.}~\bibnamefont
  {McEwen}}, \bibinfo {author} {\bibfnamefont {A.}~\bibnamefont {Megrant}},
  \bibinfo {author} {\bibfnamefont {X.}~\bibnamefont {Mi}}, \bibinfo {author}
  {\bibfnamefont {K.}~\bibnamefont {Michielsen}}, \bibinfo {author}
  {\bibfnamefont {M.}~\bibnamefont {Mohseni}}, \bibinfo {author} {\bibfnamefont
  {J.}~\bibnamefont {Mutus}}, \bibinfo {author} {\bibfnamefont
  {O.}~\bibnamefont {Naaman}}, \bibinfo {author} {\bibfnamefont
  {M.}~\bibnamefont {Neeley}}, \bibinfo {author} {\bibfnamefont
  {C.}~\bibnamefont {Neill}}, \bibinfo {author} {\bibfnamefont {M.~Y.}\
  \bibnamefont {Niu}}, \bibinfo {author} {\bibfnamefont {E.}~\bibnamefont
  {Ostby}}, \bibinfo {author} {\bibfnamefont {A.}~\bibnamefont {Petukhov}},
  \bibinfo {author} {\bibfnamefont {J.~C.}\ \bibnamefont {Platt}}, \bibinfo
  {author} {\bibfnamefont {C.}~\bibnamefont {Quintana}}, \bibinfo {author}
  {\bibfnamefont {E.~G.}\ \bibnamefont {Rieffel}}, \bibinfo {author}
  {\bibfnamefont {P.}~\bibnamefont {Roushan}}, \bibinfo {author} {\bibfnamefont
  {N.~C.}\ \bibnamefont {Rubin}}, \bibinfo {author} {\bibfnamefont
  {D.}~\bibnamefont {Sank}}, \bibinfo {author} {\bibfnamefont {K.~J.}\
  \bibnamefont {Satzinger}}, \bibinfo {author} {\bibfnamefont {V.}~\bibnamefont
  {Smelyanskiy}}, \bibinfo {author} {\bibfnamefont {K.~J.}\ \bibnamefont
  {Sung}}, \bibinfo {author} {\bibfnamefont {M.~D.}\ \bibnamefont
  {Trevithick}}, \bibinfo {author} {\bibfnamefont {A.}~\bibnamefont
  {Vainsencher}}, \bibinfo {author} {\bibfnamefont {B.}~\bibnamefont
  {Villalonga}}, \bibinfo {author} {\bibfnamefont {T.}~\bibnamefont {White}},
  \bibinfo {author} {\bibfnamefont {Z.~J.}\ \bibnamefont {Yao}}, \bibinfo
  {author} {\bibfnamefont {P.}~\bibnamefont {Yeh}}, \bibinfo {author}
  {\bibfnamefont {A.}~\bibnamefont {Zalcman}}, \bibinfo {author} {\bibfnamefont
  {H.}~\bibnamefont {Neven}}, \ and\ \bibinfo {author} {\bibfnamefont {J.~M.}\
  \bibnamefont {Martinis}},\ }\href {\doibase 10.1038/s41586-019-1666-5}
  {\bibfield  {journal} {\bibinfo  {journal} {Nature}\ }\textbf {\bibinfo
  {volume} {574}},\ \bibinfo {pages} {505} (\bibinfo {year}
  {2019})}\BibitemShut {NoStop}%
\bibitem [{\citenamefont {Braunstein}\ \emph {et~al.}(2001)\citenamefont
  {Braunstein}, \citenamefont {Fuchs}, \citenamefont {Kimble},\ and\
  \citenamefont {van Loock}}]{Braunstein2001}%
  \BibitemOpen
  \bibfield  {author} {\bibinfo {author} {\bibfnamefont {S.~L.}\ \bibnamefont
  {Braunstein}}, \bibinfo {author} {\bibfnamefont {C.~A.}\ \bibnamefont
  {Fuchs}}, \bibinfo {author} {\bibfnamefont {H.~J.}\ \bibnamefont {Kimble}}, \
  and\ \bibinfo {author} {\bibfnamefont {P.}~\bibnamefont {van Loock}},\ }\href
  {\doibase 10.1103/PhysRevA.64.022321} {\bibfield  {journal} {\bibinfo
  {journal} {Phys. Rev. A}\ }\textbf {\bibinfo {volume} {64}},\ \bibinfo
  {pages} {022321} (\bibinfo {year} {2001})}\BibitemShut {NoStop}%
\bibitem [{\citenamefont {Pirandola}\ and\ \citenamefont
  {Mancini}(2006)}]{Pirandola2006}%
  \BibitemOpen
  \bibfield  {author} {\bibinfo {author} {\bibfnamefont {S.}~\bibnamefont
  {Pirandola}}\ and\ \bibinfo {author} {\bibfnamefont {S.}~\bibnamefont
  {Mancini}},\ }\href {\doibase 10.1134/S1054660X06100057} {\bibfield
  {journal} {\bibinfo  {journal} {Laser Phys.}\ }\textbf {\bibinfo {volume}
  {16}},\ \bibinfo {pages} {1418} (\bibinfo {year} {2006})}\BibitemShut
  {NoStop}%
\bibitem [{\citenamefont {Awschalom}\ \emph {et~al.}(2021)\citenamefont
  {Awschalom}, \citenamefont {Berggren}, \citenamefont {Bernien}, \citenamefont
  {Bhave}, \citenamefont {Carr}, \citenamefont {Davids}, \citenamefont
  {Economou}, \citenamefont {Englund}, \citenamefont {Faraon}, \citenamefont
  {Fejer}, \citenamefont {Guha}, \citenamefont {Gustafsson}, \citenamefont
  {Hu}, \citenamefont {Jiang}, \citenamefont {Kim}, \citenamefont {Korzh},
  \citenamefont {Kumar}, \citenamefont {Kwiat}, \citenamefont
  {Lon\ifmmode~\check{c}\else \v{c}\fi{}ar}, \citenamefont {Lukin},
  \citenamefont {Miller}, \citenamefont {Monroe}, \citenamefont {Nam},
  \citenamefont {Narang}, \citenamefont {Orcutt}, \citenamefont {Raymer},
  \citenamefont {Safavi-Naeini}, \citenamefont {Spiropulu}, \citenamefont
  {Srinivasan}, \citenamefont {Sun}, \citenamefont {Vu\ifmmode \check{c}\else
  \v{c}\fi{}kovi\ifmmode~\acute{c}\else \'{c}\fi{}}, \citenamefont {Waks},
  \citenamefont {Walsworth}, \citenamefont {Weiner},\ and\ \citenamefont
  {Zhang}}]{Awschalom2021}%
  \BibitemOpen
  \bibfield  {author} {\bibinfo {author} {\bibfnamefont {D.}~\bibnamefont
  {Awschalom}}, \bibinfo {author} {\bibfnamefont {K.~K.}\ \bibnamefont
  {Berggren}}, \bibinfo {author} {\bibfnamefont {H.}~\bibnamefont {Bernien}},
  \bibinfo {author} {\bibfnamefont {S.}~\bibnamefont {Bhave}}, \bibinfo
  {author} {\bibfnamefont {L.~D.}\ \bibnamefont {Carr}}, \bibinfo {author}
  {\bibfnamefont {P.}~\bibnamefont {Davids}}, \bibinfo {author} {\bibfnamefont
  {S.~E.}\ \bibnamefont {Economou}}, \bibinfo {author} {\bibfnamefont
  {D.}~\bibnamefont {Englund}}, \bibinfo {author} {\bibfnamefont
  {A.}~\bibnamefont {Faraon}}, \bibinfo {author} {\bibfnamefont
  {M.}~\bibnamefont {Fejer}}, \bibinfo {author} {\bibfnamefont
  {S.}~\bibnamefont {Guha}}, \bibinfo {author} {\bibfnamefont {M.~V.}\
  \bibnamefont {Gustafsson}}, \bibinfo {author} {\bibfnamefont
  {E.}~\bibnamefont {Hu}}, \bibinfo {author} {\bibfnamefont {L.}~\bibnamefont
  {Jiang}}, \bibinfo {author} {\bibfnamefont {J.}~\bibnamefont {Kim}}, \bibinfo
  {author} {\bibfnamefont {B.}~\bibnamefont {Korzh}}, \bibinfo {author}
  {\bibfnamefont {P.}~\bibnamefont {Kumar}}, \bibinfo {author} {\bibfnamefont
  {P.~G.}\ \bibnamefont {Kwiat}}, \bibinfo {author} {\bibfnamefont
  {M.}~\bibnamefont {Lon\ifmmode~\check{c}\else \v{c}\fi{}ar}}, \bibinfo
  {author} {\bibfnamefont {M.~D.}\ \bibnamefont {Lukin}}, \bibinfo {author}
  {\bibfnamefont {D.~A.}\ \bibnamefont {Miller}}, \bibinfo {author}
  {\bibfnamefont {C.}~\bibnamefont {Monroe}}, \bibinfo {author} {\bibfnamefont
  {S.~W.}\ \bibnamefont {Nam}}, \bibinfo {author} {\bibfnamefont
  {P.}~\bibnamefont {Narang}}, \bibinfo {author} {\bibfnamefont {J.~S.}\
  \bibnamefont {Orcutt}}, \bibinfo {author} {\bibfnamefont {M.~G.}\
  \bibnamefont {Raymer}}, \bibinfo {author} {\bibfnamefont {A.~H.}\
  \bibnamefont {Safavi-Naeini}}, \bibinfo {author} {\bibfnamefont
  {M.}~\bibnamefont {Spiropulu}}, \bibinfo {author} {\bibfnamefont
  {K.}~\bibnamefont {Srinivasan}}, \bibinfo {author} {\bibfnamefont
  {S.}~\bibnamefont {Sun}}, \bibinfo {author} {\bibfnamefont {J.}~\bibnamefont
  {Vu\ifmmode \check{c}\else \v{c}\fi{}kovi\ifmmode~\acute{c}\else
  \'{c}\fi{}}}, \bibinfo {author} {\bibfnamefont {E.}~\bibnamefont {Waks}},
  \bibinfo {author} {\bibfnamefont {R.}~\bibnamefont {Walsworth}}, \bibinfo
  {author} {\bibfnamefont {A.~M.}\ \bibnamefont {Weiner}}, \ and\ \bibinfo
  {author} {\bibfnamefont {Z.}~\bibnamefont {Zhang}},\ }\href {\doibase
  10.1103/PRXQuantum.2.017002} {\bibfield  {journal} {\bibinfo  {journal} {PRX
  Quantum}\ }\textbf {\bibinfo {volume} {2}},\ \bibinfo {pages} {017002}
  (\bibinfo {year} {2021})}\BibitemShut {NoStop}%
\bibitem [{\citenamefont {Clauser}\ \emph {et~al.}(1969)\citenamefont
  {Clauser}, \citenamefont {Horne}, \citenamefont {Shimony},\ and\
  \citenamefont {Holt}}]{Clauser1969}%
  \BibitemOpen
  \bibfield  {author} {\bibinfo {author} {\bibfnamefont {J.~F.}\ \bibnamefont
  {Clauser}}, \bibinfo {author} {\bibfnamefont {M.~A.}\ \bibnamefont {Horne}},
  \bibinfo {author} {\bibfnamefont {A.}~\bibnamefont {Shimony}}, \ and\
  \bibinfo {author} {\bibfnamefont {R.~A.}\ \bibnamefont {Holt}},\ }\href
  {\doibase 10.1103/PhysRevLett.23.880} {\bibfield  {journal} {\bibinfo
  {journal} {Phys. Rev. Lett.}\ }\textbf {\bibinfo {volume} {23}},\ \bibinfo
  {pages} {880} (\bibinfo {year} {1969})}\BibitemShut {NoStop}%
\bibitem [{\citenamefont {Di~Candia}\ \emph {et~al.}(2015)\citenamefont
  {Di~Candia}, \citenamefont {Fedorov}, \citenamefont {Zhong}, \citenamefont
  {Felicetti}, \citenamefont {Menzel}, \citenamefont {Sanz}, \citenamefont
  {Deppe}, \citenamefont {Marx}, \citenamefont {Gross},\ and\ \citenamefont
  {Solano}}]{DiCandia2015}%
  \BibitemOpen
  \bibfield  {author} {\bibinfo {author} {\bibfnamefont {R.}~\bibnamefont
  {Di~Candia}}, \bibinfo {author} {\bibfnamefont {K.~G.}\ \bibnamefont
  {Fedorov}}, \bibinfo {author} {\bibfnamefont {L.}~\bibnamefont {Zhong}},
  \bibinfo {author} {\bibfnamefont {S.}~\bibnamefont {Felicetti}}, \bibinfo
  {author} {\bibfnamefont {E.~P.}\ \bibnamefont {Menzel}}, \bibinfo {author}
  {\bibfnamefont {M.}~\bibnamefont {Sanz}}, \bibinfo {author} {\bibfnamefont
  {F.}~\bibnamefont {Deppe}}, \bibinfo {author} {\bibfnamefont
  {A.}~\bibnamefont {Marx}}, \bibinfo {author} {\bibfnamefont {R.}~\bibnamefont
  {Gross}}, \ and\ \bibinfo {author} {\bibfnamefont {E.}~\bibnamefont
  {Solano}},\ }\href {\doibase 10.1140/epjqt/s40507-015-0038-9} {\bibfield
  {journal} {\bibinfo  {journal} {EPJ Quantum Technol.}\ }\textbf {\bibinfo
  {volume} {2}},\ \bibinfo {pages} {25} (\bibinfo {year} {2015})}\BibitemShut
  {NoStop}%
\bibitem [{\citenamefont {Yamamoto}\ \emph {et~al.}(2008)\citenamefont
  {Yamamoto}, \citenamefont {Inomata}, \citenamefont {Watanabe}, \citenamefont
  {Matsuba}, \citenamefont {Miyazaki}, \citenamefont {Oliver}, \citenamefont
  {Nakamura},\ and\ \citenamefont {Tsai}}]{Yamamoto2008}%
  \BibitemOpen
  \bibfield  {author} {\bibinfo {author} {\bibfnamefont {T.}~\bibnamefont
  {Yamamoto}}, \bibinfo {author} {\bibfnamefont {K.}~\bibnamefont {Inomata}},
  \bibinfo {author} {\bibfnamefont {M.}~\bibnamefont {Watanabe}}, \bibinfo
  {author} {\bibfnamefont {K.}~\bibnamefont {Matsuba}}, \bibinfo {author}
  {\bibfnamefont {T.}~\bibnamefont {Miyazaki}}, \bibinfo {author}
  {\bibfnamefont {W.~D.}\ \bibnamefont {Oliver}}, \bibinfo {author}
  {\bibfnamefont {Y.}~\bibnamefont {Nakamura}}, \ and\ \bibinfo {author}
  {\bibfnamefont {J.~S.}\ \bibnamefont {Tsai}},\ }\href {\doibase
  10.1063/1.2964182} {\bibfield  {journal} {\bibinfo  {journal} {Appl. Phys.
  Lett.}\ }\textbf {\bibinfo {volume} {93}},\ \bibinfo {pages} {042510}
  (\bibinfo {year} {2008})}\BibitemShut {NoStop}%
\bibitem [{\citenamefont {Menzel}\ \emph {et~al.}(2012)\citenamefont {Menzel},
  \citenamefont {Di~Candia}, \citenamefont {Deppe}, \citenamefont {Eder},
  \citenamefont {Zhong}, \citenamefont {Ihmig}, \citenamefont {Haeberlein},
  \citenamefont {Baust}, \citenamefont {Hoffmann}, \citenamefont {Ballester},
  \citenamefont {Inomata}, \citenamefont {Yamamoto}, \citenamefont {Nakamura},
  \citenamefont {Solano}, \citenamefont {Marx},\ and\ \citenamefont
  {Gross}}]{Menzel2012}%
  \BibitemOpen
  \bibfield  {author} {\bibinfo {author} {\bibfnamefont {E.~P.}\ \bibnamefont
  {Menzel}}, \bibinfo {author} {\bibfnamefont {R.}~\bibnamefont {Di~Candia}},
  \bibinfo {author} {\bibfnamefont {F.}~\bibnamefont {Deppe}}, \bibinfo
  {author} {\bibfnamefont {P.}~\bibnamefont {Eder}}, \bibinfo {author}
  {\bibfnamefont {L.}~\bibnamefont {Zhong}}, \bibinfo {author} {\bibfnamefont
  {M.}~\bibnamefont {Ihmig}}, \bibinfo {author} {\bibfnamefont
  {M.}~\bibnamefont {Haeberlein}}, \bibinfo {author} {\bibfnamefont
  {A.}~\bibnamefont {Baust}}, \bibinfo {author} {\bibfnamefont
  {E.}~\bibnamefont {Hoffmann}}, \bibinfo {author} {\bibfnamefont
  {D.}~\bibnamefont {Ballester}}, \bibinfo {author} {\bibfnamefont
  {K.}~\bibnamefont {Inomata}}, \bibinfo {author} {\bibfnamefont
  {T.}~\bibnamefont {Yamamoto}}, \bibinfo {author} {\bibfnamefont
  {Y.}~\bibnamefont {Nakamura}}, \bibinfo {author} {\bibfnamefont
  {E.}~\bibnamefont {Solano}}, \bibinfo {author} {\bibfnamefont
  {A.}~\bibnamefont {Marx}}, \ and\ \bibinfo {author} {\bibfnamefont
  {R.}~\bibnamefont {Gross}},\ }\href {\doibase 10.1103/PhysRevLett.109.250502}
  {\bibfield  {journal} {\bibinfo  {journal} {Phys. Rev. Lett.}\ }\textbf
  {\bibinfo {volume} {109}},\ \bibinfo {pages} {250502} (\bibinfo {year}
  {2012})}\BibitemShut {NoStop}%
\bibitem [{\citenamefont {Fedorov}\ \emph {et~al.}(2018)\citenamefont
  {Fedorov}, \citenamefont {Pogorzalek}, \citenamefont {Las~Heras},
  \citenamefont {Sanz}, \citenamefont {Yard}, \citenamefont {Eder},
  \citenamefont {Fischer}, \citenamefont {Goetz}, \citenamefont {Xie},
  \citenamefont {Inomata}, \citenamefont {Nakamura}, \citenamefont {Di~Candia},
  \citenamefont {Solano}, \citenamefont {Marx}, \citenamefont {Deppe},\ and\
  \citenamefont {Gross}}]{Fedorov2018}%
  \BibitemOpen
  \bibfield  {author} {\bibinfo {author} {\bibfnamefont {K.~G.}\ \bibnamefont
  {Fedorov}}, \bibinfo {author} {\bibfnamefont {S.}~\bibnamefont {Pogorzalek}},
  \bibinfo {author} {\bibfnamefont {U.}~\bibnamefont {Las~Heras}}, \bibinfo
  {author} {\bibfnamefont {M.}~\bibnamefont {Sanz}}, \bibinfo {author}
  {\bibfnamefont {P.}~\bibnamefont {Yard}}, \bibinfo {author} {\bibfnamefont
  {P.}~\bibnamefont {Eder}}, \bibinfo {author} {\bibfnamefont {M.}~\bibnamefont
  {Fischer}}, \bibinfo {author} {\bibfnamefont {J.}~\bibnamefont {Goetz}},
  \bibinfo {author} {\bibfnamefont {E.}~\bibnamefont {Xie}}, \bibinfo {author}
  {\bibfnamefont {K.}~\bibnamefont {Inomata}}, \bibinfo {author} {\bibfnamefont
  {Y.}~\bibnamefont {Nakamura}}, \bibinfo {author} {\bibfnamefont
  {R.}~\bibnamefont {Di~Candia}}, \bibinfo {author} {\bibfnamefont
  {E.}~\bibnamefont {Solano}}, \bibinfo {author} {\bibfnamefont
  {A.}~\bibnamefont {Marx}}, \bibinfo {author} {\bibfnamefont {F.}~\bibnamefont
  {Deppe}}, \ and\ \bibinfo {author} {\bibfnamefont {R.}~\bibnamefont
  {Gross}},\ }\href {\doibase 10.1038/s41598-018-24742-z} {\bibfield  {journal}
  {\bibinfo  {journal} {Sci. Rep.}\ }\textbf {\bibinfo {volume} {8}},\ \bibinfo
  {pages} {6416} (\bibinfo {year} {2018})}\BibitemShut {NoStop}%
\bibitem [{\citenamefont {Pogorzalek}\ \emph {et~al.}(2019)\citenamefont
  {Pogorzalek}, \citenamefont {Fedorov}, \citenamefont {Xu}, \citenamefont
  {Parra-Rodriguez}, \citenamefont {Sanz}, \citenamefont {Fischer},
  \citenamefont {Xie}, \citenamefont {Inomata}, \citenamefont {Nakamura},
  \citenamefont {Solano}, \citenamefont {Marx}, \citenamefont {Deppe},\ and\
  \citenamefont {Gross}}]{Pogorzalek2019}%
  \BibitemOpen
  \bibfield  {author} {\bibinfo {author} {\bibfnamefont {S.}~\bibnamefont
  {Pogorzalek}}, \bibinfo {author} {\bibfnamefont {K.~G.}\ \bibnamefont
  {Fedorov}}, \bibinfo {author} {\bibfnamefont {M.}~\bibnamefont {Xu}},
  \bibinfo {author} {\bibfnamefont {A.}~\bibnamefont {Parra-Rodriguez}},
  \bibinfo {author} {\bibfnamefont {M.}~\bibnamefont {Sanz}}, \bibinfo {author}
  {\bibfnamefont {M.}~\bibnamefont {Fischer}}, \bibinfo {author} {\bibfnamefont
  {E.}~\bibnamefont {Xie}}, \bibinfo {author} {\bibfnamefont {K.}~\bibnamefont
  {Inomata}}, \bibinfo {author} {\bibfnamefont {Y.}~\bibnamefont {Nakamura}},
  \bibinfo {author} {\bibfnamefont {E.}~\bibnamefont {Solano}}, \bibinfo
  {author} {\bibfnamefont {A.}~\bibnamefont {Marx}}, \bibinfo {author}
  {\bibfnamefont {F.}~\bibnamefont {Deppe}}, \ and\ \bibinfo {author}
  {\bibfnamefont {R.}~\bibnamefont {Gross}},\ }\href {\doibase
  10.1038/s41467-019-10727-7} {\bibfield  {journal} {\bibinfo  {journal} {Nat.
  Commun.}\ }\textbf {\bibinfo {volume} {10}},\ \bibinfo {pages} {2604}
  (\bibinfo {year} {2019})}\BibitemShut {NoStop}%
\bibitem [{Sup()}]{SupMat}%
  \BibitemOpen
  \href@noop {} {\bibinfo  {journal} {Supplemental Material}\ }\BibitemShut
  {NoStop}%
\bibitem [{\citenamefont {Grosshans}\ and\ \citenamefont
  {Grangier}(2002)}]{Grosshans2002}%
  \BibitemOpen
\bibfield  {journal} {  }\bibfield  {author} {\bibinfo {author} {\bibfnamefont
  {F.}~\bibnamefont {Grosshans}}\ and\ \bibinfo {author} {\bibfnamefont
  {P.}~\bibnamefont {Grangier}},\ }\href {\doibase
  10.1103/PhysRevLett.88.057902} {\bibfield  {journal} {\bibinfo  {journal}
  {Phys. Rev. Lett.}\ }\textbf {\bibinfo {volume} {88}},\ \bibinfo {pages}
  {057902} (\bibinfo {year} {2002})}\BibitemShut {NoStop}%
\bibitem [{\citenamefont {Sivak}\ \emph {et~al.}(2020)\citenamefont {Sivak},
  \citenamefont {Shankar}, \citenamefont {Liu}, \citenamefont {Aumentado},\
  and\ \citenamefont {Devoret}}]{Sivak2020}%
  \BibitemOpen
  \bibfield  {author} {\bibinfo {author} {\bibfnamefont {V.~V.}\ \bibnamefont
  {Sivak}}, \bibinfo {author} {\bibfnamefont {S.}~\bibnamefont {Shankar}},
  \bibinfo {author} {\bibfnamefont {G.}~\bibnamefont {Liu}}, \bibinfo {author}
  {\bibfnamefont {J.}~\bibnamefont {Aumentado}}, \ and\ \bibinfo {author}
  {\bibfnamefont {M.~H.}\ \bibnamefont {Devoret}},\ }\href {\doibase
  10.1103/PhysRevApplied.13.024014} {\bibfield  {journal} {\bibinfo  {journal}
  {Phys. Rev. Applied}\ }\textbf {\bibinfo {volume} {13}},\ \bibinfo {pages}
  {024014} (\bibinfo {year} {2020})}\BibitemShut {NoStop}%
\bibitem [{\citenamefont {Renger}\ \emph {et~al.}(2020)\citenamefont {Renger},
  \citenamefont {Pogorzalek}, \citenamefont {Chen}, \citenamefont {Nojiri},
  \citenamefont {Inomata}, \citenamefont {Nakamura}, \citenamefont {Partanen},
  \citenamefont {Marx}, \citenamefont {Gross}, \citenamefont {Deppe},\ and\
  \citenamefont {Fedorov}}]{Renger2020}%
  \BibitemOpen
  \bibfield  {author} {\bibinfo {author} {\bibfnamefont {M.}~\bibnamefont
  {Renger}}, \bibinfo {author} {\bibfnamefont {S.}~\bibnamefont {Pogorzalek}},
  \bibinfo {author} {\bibfnamefont {Q.}~\bibnamefont {Chen}}, \bibinfo {author}
  {\bibfnamefont {Y.}~\bibnamefont {Nojiri}}, \bibinfo {author} {\bibfnamefont
  {K.}~\bibnamefont {Inomata}}, \bibinfo {author} {\bibfnamefont
  {Y.}~\bibnamefont {Nakamura}}, \bibinfo {author} {\bibfnamefont
  {M.}~\bibnamefont {Partanen}}, \bibinfo {author} {\bibfnamefont
  {A.}~\bibnamefont {Marx}}, \bibinfo {author} {\bibfnamefont {R.}~\bibnamefont
  {Gross}}, \bibinfo {author} {\bibfnamefont {F.}~\bibnamefont {Deppe}}, \ and\
  \bibinfo {author} {\bibfnamefont {K.~G.}\ \bibnamefont {Fedorov}},\
  }\href@noop {} {\  (\bibinfo {year} {2020})},\ \Eprint
  {http://arxiv.org/abs/2011.00914} {arXiv:2011.00914 [quant-ph]} \BibitemShut
  {NoStop}%
\bibitem [{\citenamefont {Liuzzo-Scorpo}\ \emph {et~al.}(2017)\citenamefont
  {Liuzzo-Scorpo}, \citenamefont {Mari}, \citenamefont {Giovannetti},\ and\
  \citenamefont {Adesso}}]{Scorpo2017}%
  \BibitemOpen
  \bibfield  {author} {\bibinfo {author} {\bibfnamefont {P.}~\bibnamefont
  {Liuzzo-Scorpo}}, \bibinfo {author} {\bibfnamefont {A.}~\bibnamefont {Mari}},
  \bibinfo {author} {\bibfnamefont {V.}~\bibnamefont {Giovannetti}}, \ and\
  \bibinfo {author} {\bibfnamefont {G.}~\bibnamefont {Adesso}},\ }\href
  {\doibase 10.1103/PhysRevLett.119.120503} {\bibfield  {journal} {\bibinfo
  {journal} {Phys. Rev. Lett.}\ }\textbf {\bibinfo {volume} {119}},\ \bibinfo
  {pages} {120503} (\bibinfo {year} {2017})}\BibitemShut {NoStop}%
\bibitem [{\citenamefont {Braunstein}\ and\ \citenamefont
  {Caves}(1994)}]{Braunstein1994}%
  \BibitemOpen
  \bibfield  {author} {\bibinfo {author} {\bibfnamefont {S.~L.}\ \bibnamefont
  {Braunstein}}\ and\ \bibinfo {author} {\bibfnamefont {C.~M.}\ \bibnamefont
  {Caves}},\ }\href {\doibase 10.1103/PhysRevLett.72.3439} {\bibfield
  {journal} {\bibinfo  {journal} {Phys. Rev. Lett.}\ }\textbf {\bibinfo
  {volume} {72}},\ \bibinfo {pages} {3439} (\bibinfo {year}
  {1994})}\BibitemShut {NoStop}%
\bibitem [{\citenamefont {Pinel}\ \emph {et~al.}(2013)\citenamefont {Pinel},
  \citenamefont {Jian}, \citenamefont {Treps}, \citenamefont {Fabre},\ and\
  \citenamefont {Braun}}]{Pinel2013}%
  \BibitemOpen
  \bibfield  {author} {\bibinfo {author} {\bibfnamefont {O.}~\bibnamefont
  {Pinel}}, \bibinfo {author} {\bibfnamefont {P.}~\bibnamefont {Jian}},
  \bibinfo {author} {\bibfnamefont {N.}~\bibnamefont {Treps}}, \bibinfo
  {author} {\bibfnamefont {C.}~\bibnamefont {Fabre}}, \ and\ \bibinfo {author}
  {\bibfnamefont {D.}~\bibnamefont {Braun}},\ }\href {\doibase
  10.1103/PhysRevA.88.040102} {\bibfield  {journal} {\bibinfo  {journal} {Phys.
  Rev. A}\ }\textbf {\bibinfo {volume} {88}},\ \bibinfo {pages} {040102}
  (\bibinfo {year} {2013})}\BibitemShut {NoStop}%
\bibitem [{\citenamefont {Xie}\ \emph {et~al.}(2018)\citenamefont {Xie},
  \citenamefont {Deppe}, \citenamefont {Renger}, \citenamefont {Repp},
  \citenamefont {Eder}, \citenamefont {Fischer}, \citenamefont {Goetz},
  \citenamefont {Pogorzalek}, \citenamefont {Fedorov}, \citenamefont {Marx},\
  and\ \citenamefont {Gross}}]{Edwar2018}%
  \BibitemOpen
  \bibfield  {author} {\bibinfo {author} {\bibfnamefont {E.}~\bibnamefont
  {Xie}}, \bibinfo {author} {\bibfnamefont {F.}~\bibnamefont {Deppe}}, \bibinfo
  {author} {\bibfnamefont {M.}~\bibnamefont {Renger}}, \bibinfo {author}
  {\bibfnamefont {D.}~\bibnamefont {Repp}}, \bibinfo {author} {\bibfnamefont
  {P.}~\bibnamefont {Eder}}, \bibinfo {author} {\bibfnamefont {M.}~\bibnamefont
  {Fischer}}, \bibinfo {author} {\bibfnamefont {J.}~\bibnamefont {Goetz}},
  \bibinfo {author} {\bibfnamefont {S.}~\bibnamefont {Pogorzalek}}, \bibinfo
  {author} {\bibfnamefont {K.~G.}\ \bibnamefont {Fedorov}}, \bibinfo {author}
  {\bibfnamefont {A.}~\bibnamefont {Marx}}, \ and\ \bibinfo {author}
  {\bibfnamefont {R.}~\bibnamefont {Gross}},\ }\href {\doibase
  10.1063/1.5029514} {\bibfield  {journal} {\bibinfo  {journal} {Appl. Phys.
  Lett.}\ }\textbf {\bibinfo {volume} {112}},\ \bibinfo {pages} {202601}
  (\bibinfo {year} {2018})}\BibitemShut {NoStop}%
\bibitem [{\citenamefont {Gao}\ \emph {et~al.}(2018)\citenamefont {Gao},
  \citenamefont {Lester}, \citenamefont {Zhang}, \citenamefont {Wang},
  \citenamefont {Rosenblum}, \citenamefont {Frunzio}, \citenamefont {Jiang},
  \citenamefont {Girvin},\ and\ \citenamefont {Schoelkopf}}]{Gao2018}%
  \BibitemOpen
  \bibfield  {author} {\bibinfo {author} {\bibfnamefont {Y.~Y.}\ \bibnamefont
  {Gao}}, \bibinfo {author} {\bibfnamefont {B.~J.}\ \bibnamefont {Lester}},
  \bibinfo {author} {\bibfnamefont {Y.}~\bibnamefont {Zhang}}, \bibinfo
  {author} {\bibfnamefont {C.}~\bibnamefont {Wang}}, \bibinfo {author}
  {\bibfnamefont {S.}~\bibnamefont {Rosenblum}}, \bibinfo {author}
  {\bibfnamefont {L.}~\bibnamefont {Frunzio}}, \bibinfo {author} {\bibfnamefont
  {L.}~\bibnamefont {Jiang}}, \bibinfo {author} {\bibfnamefont {S.~M.}\
  \bibnamefont {Girvin}}, \ and\ \bibinfo {author} {\bibfnamefont {R.~J.}\
  \bibnamefont {Schoelkopf}},\ }\href {\doibase 10.1103/PhysRevX.8.021073}
  {\bibfield  {journal} {\bibinfo  {journal} {Phys. Rev. X}\ }\textbf {\bibinfo
  {volume} {8}},\ \bibinfo {pages} {021073} (\bibinfo {year}
  {2018})}\BibitemShut {NoStop}%
\bibitem [{\citenamefont {Mirhosseini}\ \emph {et~al.}(2020)\citenamefont
  {Mirhosseini}, \citenamefont {Sipahigil}, \citenamefont {Kalaee},\ and\
  \citenamefont {Painter}}]{Mirhosseini2020}%
  \BibitemOpen
  \bibfield  {author} {\bibinfo {author} {\bibfnamefont {M.}~\bibnamefont
  {Mirhosseini}}, \bibinfo {author} {\bibfnamefont {A.}~\bibnamefont
  {Sipahigil}}, \bibinfo {author} {\bibfnamefont {M.}~\bibnamefont {Kalaee}}, \
  and\ \bibinfo {author} {\bibfnamefont {O.}~\bibnamefont {Painter}},\ }\href
  {\doibase 10.1038/s41586-020-3038-6} {\bibfield  {journal} {\bibinfo
  {journal} {Nature}\ }\textbf {\bibinfo {volume} {588}},\ \bibinfo {pages}
  {599} (\bibinfo {year} {2020})}\BibitemShut {NoStop}%
\bibitem [{\citenamefont {Fedorov}\ \emph {et~al.}(2016)\citenamefont
  {Fedorov}, \citenamefont {Zhong}, \citenamefont {Pogorzalek}, \citenamefont
  {Eder}, \citenamefont {Fischer}, \citenamefont {Goetz}, \citenamefont {Xie},
  \citenamefont {Wulschner}, \citenamefont {Inomata}, \citenamefont {Yamamoto},
  \citenamefont {Nakamura}, \citenamefont {{Di Candia}}, \citenamefont {{Las
  Heras}}, \citenamefont {Sanz}, \citenamefont {Solano}, \citenamefont
  {Menzel}, \citenamefont {Deppe}, \citenamefont {Marx},\ and\ \citenamefont
  {Gross}}]{Fedorov2016}%
  \BibitemOpen
  \bibfield  {author} {\bibinfo {author} {\bibfnamefont {K.~G.}\ \bibnamefont
  {Fedorov}}, \bibinfo {author} {\bibfnamefont {L.}~\bibnamefont {Zhong}},
  \bibinfo {author} {\bibfnamefont {S.}~\bibnamefont {Pogorzalek}}, \bibinfo
  {author} {\bibfnamefont {P.}~\bibnamefont {Eder}}, \bibinfo {author}
  {\bibfnamefont {M.}~\bibnamefont {Fischer}}, \bibinfo {author} {\bibfnamefont
  {J.}~\bibnamefont {Goetz}}, \bibinfo {author} {\bibfnamefont
  {E.}~\bibnamefont {Xie}}, \bibinfo {author} {\bibfnamefont {F.}~\bibnamefont
  {Wulschner}}, \bibinfo {author} {\bibfnamefont {K.}~\bibnamefont {Inomata}},
  \bibinfo {author} {\bibfnamefont {T.}~\bibnamefont {Yamamoto}}, \bibinfo
  {author} {\bibfnamefont {Y.}~\bibnamefont {Nakamura}}, \bibinfo {author}
  {\bibfnamefont {R.}~\bibnamefont {{Di Candia}}}, \bibinfo {author}
  {\bibfnamefont {U.}~\bibnamefont {{Las Heras}}}, \bibinfo {author}
  {\bibfnamefont {M.}~\bibnamefont {Sanz}}, \bibinfo {author} {\bibfnamefont
  {E.}~\bibnamefont {Solano}}, \bibinfo {author} {\bibfnamefont {E.~P.}\
  \bibnamefont {Menzel}}, \bibinfo {author} {\bibfnamefont {F.}~\bibnamefont
  {Deppe}}, \bibinfo {author} {\bibfnamefont {A.}~\bibnamefont {Marx}}, \ and\
  \bibinfo {author} {\bibfnamefont {R.}~\bibnamefont {Gross}},\ }\href
  {\doibase 10.1103/PhysRevLett.117.020502} {\bibfield  {journal} {\bibinfo
  {journal} {Phys. Rev. Lett.}\ }\textbf {\bibinfo {volume} {117}},\ \bibinfo
  {pages} {020502} (\bibinfo {year} {2016})}\BibitemShut {NoStop}%
\bibitem [{\citenamefont {Eichler}\ \emph {et~al.}(2011)\citenamefont
  {Eichler}, \citenamefont {Bozyigit}, \citenamefont {Lang}, \citenamefont
  {Baur}, \citenamefont {Steffen}, \citenamefont {Fink}, \citenamefont
  {Filipp},\ and\ \citenamefont {Wallraff}}]{Eichler2011b}%
  \BibitemOpen
  \bibfield  {author} {\bibinfo {author} {\bibfnamefont {C.}~\bibnamefont
  {Eichler}}, \bibinfo {author} {\bibfnamefont {D.}~\bibnamefont {Bozyigit}},
  \bibinfo {author} {\bibfnamefont {C.}~\bibnamefont {Lang}}, \bibinfo {author}
  {\bibfnamefont {M.}~\bibnamefont {Baur}}, \bibinfo {author} {\bibfnamefont
  {L.}~\bibnamefont {Steffen}}, \bibinfo {author} {\bibfnamefont {J.~M.}\
  \bibnamefont {Fink}}, \bibinfo {author} {\bibfnamefont {S.}~\bibnamefont
  {Filipp}}, \ and\ \bibinfo {author} {\bibfnamefont {A.}~\bibnamefont
  {Wallraff}},\ }\href {\doibase 10.1103/PhysRevLett.107.113601} {\bibfield
  {journal} {\bibinfo  {journal} {Phys. Rev. Lett.}\ }\textbf {\bibinfo
  {volume} {107}},\ \bibinfo {pages} {113601} (\bibinfo {year}
  {2011})}\BibitemShut {NoStop}%
\end{thebibliography}%

\end{document}